\documentclass{aastex}
\usepackage{spr-astr-addons}
\usepackage{url}

\RequirePackage{color}

\newcommand{\emaila}{psotomayor@iar.unlp.edu.ar}
\newcommand{\grad}{$^{\circ}$}

\begin{document}




\title{Equatorial outflows driven by jets in Population III microquasars}
\slugcomment{Not to appear in Nonlearned J., 45.}
\shorttitle{Equatorial outflows from MQs}
\shortauthors{Sotomayor~Checa et al.}

\author{Pablo Sotomayor Checa\altaffilmark{1,2}}
\affil{\emaila}
\and \author{Gustavo E. Romero\altaffilmark{1,2}} 
\and \author{Valenti Bosch-Ramon\altaffilmark{3}}

\altaffiltext{1}{Instituto Argentino de Radioastronom\'{\i}a (CONICET; CICPBA), C.C. No. 5, 1894 Villa Elisa, Argentina.}
\altaffiltext{2}{Facultad de Ciencias Astron\'omicas y Geof\'{\i}sicas, Universidad Nacional de La Plata, Paseo del Bosque s/n, 1900, La Plata, Argentina.}
\altaffiltext{3}{Departament de F\'{i}sica Qu\`antica i Astrof\'{i}sica, Institut de Ci\`encies del Cosmos (ICC), Universitat de Barcelona (IEEC-UB), Mart\'{i} i Franqu\`es 1, E08028 Barcelona, Spain.}


\begin{abstract}
Binary systems of Population III can evolve to microquasars when one of the stars collapses into a black hole. When the compact object accretes matter at a rate greater than the Eddington rate, powerful jets and winds driven by strong radiation pressure should form. We investigate the structure of the jet-wind system for a model of Population III microquasar on scales beyond the jet-wind formation region. Using relativistic hydrodynamic simulations we find that the ratio of kinetic power between the jet and the disk wind determines the configuration of the system. When the power is dominated by the wind, the jet fills a narrow channel, collimated by the dense outflow. When the jet dominates the power of the system, part of its energy is diverted turning the wind into a quasi-equatorial flow, while the jet widens. 
From the results of our simulations, we implement semi-analytical calculations of the impact of the quasi-equatorial wind on scales of the order of the size of the binary system. Our results indicate that Population III microquasars might inject gamma rays and relativistic particles into the early intergalactic medium, contributing to its reionization at large distances from the binary system. 
\end{abstract}

\section{Introduction}
\label{sec:intro}
The accretion of matter onto compact objects powers a wide variety of astrophysical phenomena, from X-ray binaries to active galactic nuclei.  When the accreting material has significant angular momentum, an accretion disk is formed around the gravitating center. The structure and spectrum of such disks depend essentially on the accretion rate $\dot{M}$. Three basic kinds of accretion regimes can be differentiated depending on how the actual accretion rate relates to the {\it critical rate} defined as,
\begin{equation}
  \dot{M}_{\rm crit} \equiv \eta \dot{M}_{\rm Edd} = \frac{L_{\rm Edd}}{c^2} \approx 1.4\times 10^{17} \frac{M}{M_{\odot}}\;\; {\rm g \;s}^{-1},
  \end{equation}
where 
\begin{equation}
  L_{\rm Edd}=\frac{4 \pi G M m_{\rm p} c}{\sigma _{\rm T}}\approx 1.3\times10^{38} \frac{M}{M_{\odot}}\;\;{\rm erg\; s}^{-1}
\end{equation} 
is the Eddington luminosity, $M$ is the mass of the accreting object, $\eta\approx 0.1$ is the radiative efficiency, and the rest of the symbols have their usual meaning. At sub-critical rates $\dot{M}\lesssim \dot{M}_{\rm crit}$ the disk is thin and can be described using the standard model developed by  \cite{shakura1973}. At very low accretion rates,  $\dot{M}<< \dot{M}_{\rm crit}$, optically-thin advection-dominated regimes are possible \citep[e.g.][]{narayan1994}. Finally, if the regime is super-critical, $\dot{M}>> \dot{M}_{\rm crit}$, the disk becomes optically-thick, geometrically slim or thick, and advection-dominated \citep[e.g.][]{paczynsky1980,abramowicz1980,calvani1981,wiita1982,beloborodov1998,fukue2000}.\par 

It was already realized by \cite{shakura1973} that super-critical disks should expel a large amount of the accreting mass as optically thick winds. The generation, evolution, and appearance of such winds have been investigated by several authors \citep[e.g.][]{meier1979,fukue2009,fukue2011,kitabatake2002}. If a strong magnetic field is present in the inner part of the disk, super-critical sources can also launch well-collimated relativistic jets \citep{eggum1985,meier2005,fukue2005,okuda2005}. The resulting jet-wind interactions can be an important ingredient of the overall structure of the system. The wind is likely present at earlier stages of the evolution of the source, so the launching of the jet could even be inhibited if the wind is sufficiently dense. If the jet escapes, then wind collimation effects can also be significant \citep[see][]{globus2016}. An extremely powerful jet, on the other hand, might push sidewards the wind producing an equatorial outflow. This could be the case of the well-known super-accreting microquasar SS433, as the existence of an equatorial wind in SS433 has been revealed by high-resolution radio interferometric observations \citep{blundell2001,paragi2002} and the source has also powerful, precessing jets with hadronic content.\par

Another scenario where jet-wind interactions in super-critical sources might be crucial is that of Population~III microquasars (PopIII~MQs) \citep{romero2018}. In such binaries mass transfer from the very massive star to the black hole occurs through overflow of the Roche lobe, since PopIII~MQs have zero metallicity and no stellar winds. This leads to accretion rates that can exceed the Eddington rate by a factor of several thousands. These sources, then, are expected to be extreme super-critical accretors with very dense winds. The determination of how winds and jets interact in these objects is important because the resulting structure can present rather different properties depending on the jet and wind power relation. For instance, a narrow jet strongly collimated by a powerful wind may not be able to keep its integrity for long because of the growth of hydrodynamical instabilities. On the other hand, a broad jet and a quasi-equatorial wind could also lead to specific consequences because of the different impact on the ambient medium; note that jets in PopIII~MQs are expected to play a role in the cosmic re-ionization after the formation of the first stars \citep{sotomayor2019}.\par 

In this paper we investigate the jet-wind interactions in super-critical microquasars of Population III using a combination of analytical models and numerical simulations. Our goal is to characterize the range of physical possibilities emerging from such interactions in systems with accretion rates largely exceeding the Eddington rate. In particular, we are interested in the formation of a quasi-equatorial outflow. \par 
The structure of this paper is the following: in the next section we present the super-critical accretion disk models and their associated winds. Then, in Sect.~\ref{sect:Jet}, we outline the jet model we adopt, whereas we devote Sect.~\ref{sect:int} to the description of the simulations of the interactions and the results. From these findings, in Sect.~\ref{sect:PopIII}, we investigate the impact of an equatorial outflow on its environment. We close with a summary and some conclusions in Sect.~\ref{sect:concl}.

\section{Super-critical accretion disks and their winds}\label{sect:accretion}

The distinctive feature of super-critical accretion regime is photon-trapping in the disk \citep[see e.g.][]{,ohsuga2003,ohsuga2005}. When the photon diffusion time-scale exceeds the accretion time-scale, photons are advected towards the compact object without being able to go out from the surface. Advection of photons partially attenuates the disk luminosity. Following \cite{narayan1994}, the photon-trapping can be parameterized considering that advection heating is a fraction of the viscous heating, $Q_{\mathrm{adv}} = Q_{\mathrm{vis}} - Q_{\mathrm{rad}} = f Q_{\mathrm{vis}}$, where $Q_{\mathrm{rad}}$ is the radiative cooling rate, and $f$ is the advection parameter. In this paper, we assumed $f$ to be constant along the entire disk.

Powerful winds are expelled from the disk by the intense radiation pressure. We consider $\dot{M}>>\dot{M}_{\rm crit}$, so practically all the accreted matter is ejected through the winds \citep{fukue2000,fukue2004,lipunova1999}. A semi-analytical study of the physical properties of the super-critical accretion disk was developed by \cite{fukue2004} in the context of the \emph{critical accretion disk} model. In that model, it is assumed that the magnetization of the disk is weak enough to have no impact on the dynamics of the accreted fluid. However, it is expected that a toroidal magnetic field will be generated in the disk (for example, through the Biermann battery mechanism). The dynamics of the accreted fluid considering the effect of the toroidal fields is discussed in \cite{akizuki2006}. Accretion disks with toroidal magnetic fields launch winds but not jets, since the latter require a large-scale poloidal component of the field \citep{beckwith2008}.

The observational appearance of highly optically thick winds from super-critical sources has been examined by \cite{nishiyama2007}. In these outflows, the apparent photosphere (surface where the optical depth measured by an observer at infinity becomes unity) can extend up to thousands of times the gravitational radius of the central relativistic object. The wind provides an additional source of ultraviolet radiation and soft X-rays. \cite{sotomayor2019} showed that in the context of Population III microquasars, the emission from disk wind may be higher than the blackbody emission from the disk.


We consider that the initial velocity of the wind is a factor of the escape velocity (similar to stellar winds from massive stars) and neglect the self-gravitating effects of the disk. The mass-loss rate into the wind, per unit surface, is given by \citep[see][]{fukue2011}:

\begin{equation}
\rho v = \frac{1}{2\pi r}\frac{{\rm d}\dot{M}}{{\rm d}r}=\frac{\dot{M}_{*}}{2\pi r_{\rm cr}}\frac{1}{r},
\end{equation}
where $r_{\rm cr}$ is the critical radius defined in the Eq. (3) of \cite{fukue2004}\footnote{In the super-critical accretion regime, there exists some critical radius $r_{\rm cr}$ such that outside $r_{\rm cr}$, the accretion rate is constant and the disk is a radiation-pressure dominated standard disk. Inside
$r_{\rm cr}$, the accretion rate decreases with the radius as to maintain the critical rate, expelling the excess of mass by the radiation-driven wind \citep{fukue2004}.}. The critical radius is obtained by equating the radiation force and the vertical component of gravity (other possible forces on the plasma are neglected). The differential mass-loss rate does not depend on $r$: $d\dot{M}(r)/dr=\dot{M}_{*}/r_{\rm cr}$ \citep[see][]{fukue2004}.

\begin{figure}[h!]
\centering
\includegraphics[scale=.25]{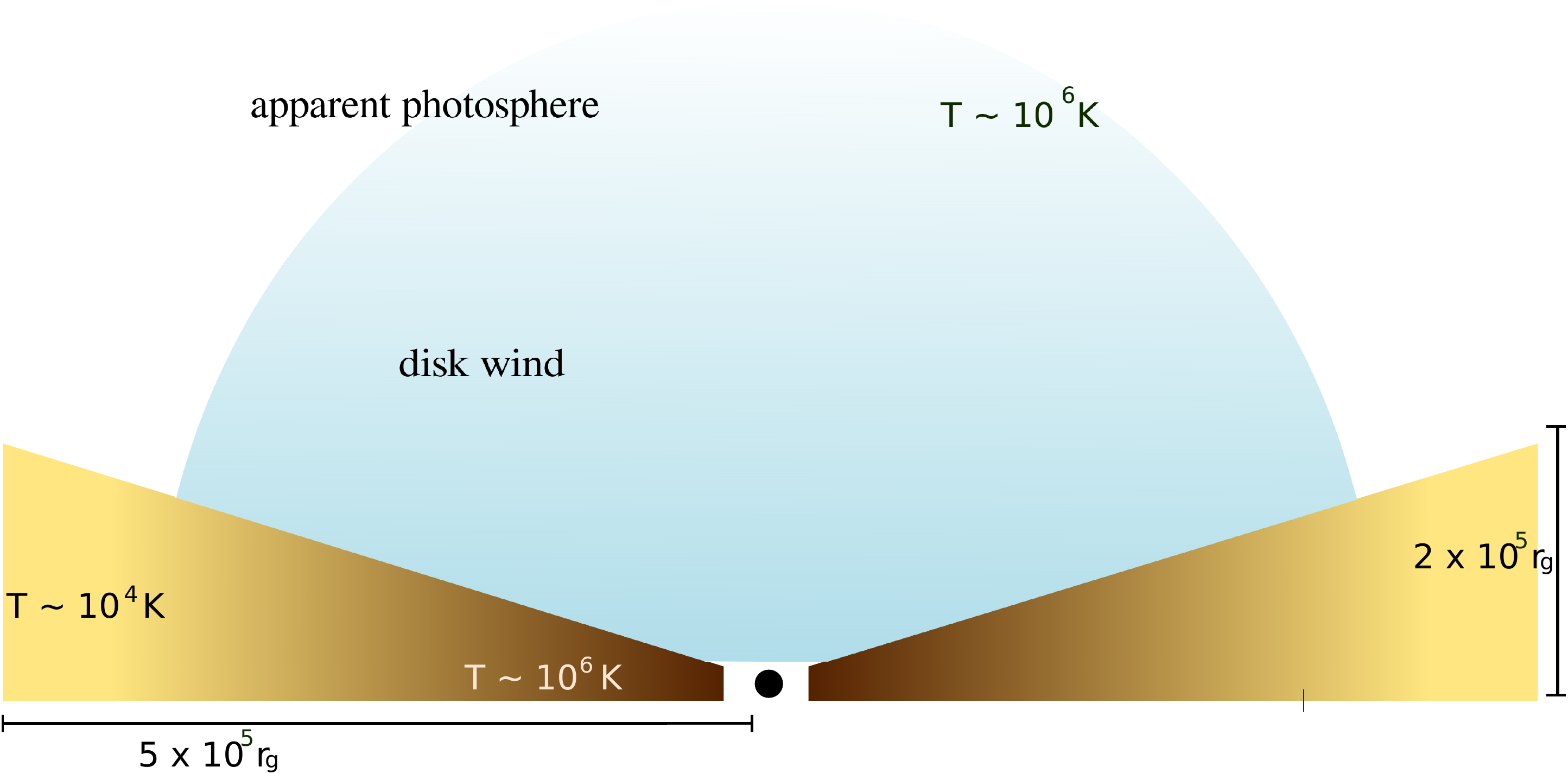}
\caption{Scheme of the considered model of accretion disk and winds in super-Eddington microquasars. Toroidal magnetic fields are expected in the disk. As long as there is no poloidal magnetic field, there is no launching of jets. The figure corresponds to the parameters adopted in Sotomayor Checa \& Romero (2019).}
\label{fig:wind}
\end{figure}

In Table~\ref{table:Parameters} we show the parameters in the different models we are going to adopt to compute the various models. We note that the jet initial radius has been adjusted to comply with resolution limitations (see Sect.~\ref{sim}), although the mass, momentum, and energy dominance of the wind external regions leaves some freedom to set the value of this parameter. We have estimated the parameters of the disk using the model developed by \cite{sotomayor2019}. The disk is supposed to be in the optically-thick advection-dominated
state with toroidal magnetic fields. The viscosity parameter is $\alpha = 0.01$, constant along the entire disk. This is consistent with models of thick accretion disks in hydrostatic equilibrium \citep{jaroszynski1980} . The magnetic field is relatively strong, $\beta = p_{\rm gas}/p_{\rm mag} = 5$, and spread throughout the disk. Detailed formulae for the most relevant physical properties of the accretion disk (thickness, temperature, magnetic field,  velocity and spectral energy distribution of the emitted radiation) can be found in \cite{akizuki2006}. A sketch of typical super-critical accretion disk with winds is shown in Fig. \ref{fig:wind}.

\begin{table*}[t]
\centering \caption{\small Parameters of the models discussed in the text}
\label{table:Parameters}      
\begin{tabular}{@{} ll}        
\hline              
Parameter [Unit] & Value\\    
\hline                   
   $M_{*}$: Donor star mass [$M_{\odot}$] & $41$ \\     
   $M_{\mathrm{BH}}$: Black hole mass [$M_{\odot}$]  & $34$ \\
   $a$: Orbital semiaxis $[R_{\odot}]$ & $36$ \\
   $\dot{M}_{\mathrm{*}}$: Mass-loss rate [$M_{\odot}\,\mathrm{yr}^{-1}$]  & $7.5\times10^{-5}$ \\
   $R_{*}$: Stellar radius $[R_{\odot}]$ & $14.2$ \\
   $r_{\mathrm{g}}$: Gravitational radius $[\mathrm{km}]$ & $50$ \\
   $\alpha$: Disk viscosity parameter & $0.01$  \\ 
   $f$: Advection parameter & $0.5$  \\
   $\beta$: Beta factor of the plasma & $5$  \\
   $r_{\rm cr}$: Critical radius $[r_{\mathrm{g}}]$ & $1.3\times 10^{4}$\\
   $L_{\rm disk}$: Disk luminosity $[\mathrm{erg\,s^{-1}}]$ & $10^{40}$\\
   $\chi$: Jet semi-opening angle tangent & $0.1$ \\
   $r_{\mathrm{0}}$: Jet initial radius $[r_{\mathrm{g}}]$ & $500$ \\
   $\dot{M}_{\mathrm{wind}}$: Total wind mass-loss rate $[M_{\odot}\,\mathrm{yr}^{-1}]$ & $7.3\times10^{-5}$ \\     
   $\dot{M}_{\mathrm{jet}}$: Total jet mass-loss rate $[M_{\odot}\,\mathrm{yr}^{-1}]$ & $3.5\times10^{-7}$ \\   
      \hline          
   Case 1: Fast jet\\
   \hline
  $v_{\mathrm{wind}}$: Wind velocity & $\sqrt{2GM/r}$  \\      
  $L_{\mathrm{wind}}$: Wind kinetic power $[\mathrm{erg\,s^{-1}}]$ & $10^{39}$ \\        
  $\Gamma_{\mathrm{jet}}$: Jet bulk Lorentz factor & $9$ \\
  $L_{\mathrm{jet}}$: Jet kinetic power $[\mathrm{erg\,s^{-1}}]$ & $10^{41}$ \\           
   \hline   
   Case 2: Heavy jet\\
   \hline  
  $v_{\mathrm{wind}}$: Wind velocity & $3\times \sqrt{2GM/r}$ \\      
  $L_{\mathrm{wind}}$: Wind kinetic power $[\mathrm{erg\,s^{-1}}]$ & $10^{40}$ \\        
  $\Gamma_{\mathrm{jet}}$: Jet bulk Lorentz factor & $1.35$ \\
  $L_{\mathrm{jet}}$: Jet kinetic power $[\mathrm{erg\,s^{-1}}]$ & $10^{40}$ \\           
 \hline
 Case 3: Very powerful jet \\
 \hline   
  $v_{\mathrm{wind}}$: Wind velocity & $9\times \sqrt{2GM/r}$ \\      
  $L_{\mathrm{wind}}$: Wind kinetic power $[\mathrm{erg\,s^{-1}}]$ & $10^{41}$ \\        
  $\Gamma_{\mathrm{jet}}$: Jet bulk Lorentz factor & $1.035$ \\
  $L_{\mathrm{jet}}$: Jet kinetic power $[\mathrm{erg\,s^{-1}}]$ & $10^{39}$ \\           
 \hline
\end{tabular}
\end{table*}

\section{Jet model}
\label{sect:Jet}

If a large-scale poloidal magnetic field develops, highly collimated, oppositely directed relativistic jets are launched from the ergosphere of the black hole \citep[see e.g.][]{tchekhovskoy2015}. Jets in super-Eddington microquasars should drill the wind to escape. The matter in the jet comes from the accreted fluid in the disk, being injected first as neutral particles \citep{romero2020}. Although the wind can load the jet, we consider that the walls of the jet are impermeable (which is a good approximation for low pressures). In all cases, we assume a conical jet that it is ejected by conversion of magnetic energy into kinetic energy in a region close to the compact object, within the disk wind.\par 

The magnetic field in the disk is initially only toroidal. A poloidal magnetic flux can be generated in situ through a turbulent dynamo powered by magnetorotational instabilities \citep{liska2018}. Then, the disk becomes a magnetically arrested disk or MAD in its inner region \citep{narayan2003}, and powerful jets are launched with Lorentz factor $\Gamma_{\rm jet} \approx 10$ \citep{sotomayor2019}. For a full discussion of the physics of these jets see e.g. \cite{romero2008}, \cite{romero2017} and \cite{romero2020}. Here we shall consider that the jets are perpendicular to the orbital plane and they do not precess. We shall investigate 3 different cases of jets. Case 1 is a fast jet launched by the black hole ergosphere. In Cases 2 and 3, we consider slower jets. These jets are similar to the case of SS433, the only known super-accreting galactic microquasar. Particularly, in Case 3 we adopt the Lorentz factor of the approaching jet of SS433. Finally, we consider that the jet is perpendicular to the orbital plane. All parameters of these models and the assumed values are shown in Table~\ref{table:Parameters}.  

\section{Jet-wind interaction}
\label{sect:int}

We implement numerical simulations to explore the physics of the interaction between the jet and the wind of the accretion disk in super-Eddington microquasars. For PopIII~MQs, we consider different values of the jet Lorentz factor, fixing the mass-injection rate. The lower the Lorentz factor is, the lower results the kinetic luminosity of the jet (see Table~\ref{table:Parameters}). Also, varying the wind velocity we set the kinetic power of the wind. In Case 1, we consider that the wind velocity is given by the escape velocity $v_{\rm wind}(r) = v_{\rm esc} (r) = \sqrt{2GM/r}$. This implies  $L_{\rm wind}\approx 10^{39}\,{\rm erg\,s^{-1}}$ for $\dot{M}_{\mathrm{*}}=7.5\times 10^{-5}\,{\rm M_\odot\,yr^{-1}}$. In Cases 2 and 3, we adopt wind velocities higher than the escape velocity, resulting in $L_{\rm wind}\approx 10^{40}\,{\rm erg\,s^{-1}}$ for $v_{\rm wind}(r) = 3\,v_{\rm esc} (r)$, and $L_{\rm wind}\approx 10^{41}\,{\rm erg\,s^{-1}}$ for $v_{\rm wind}(r) = 9\,v_{\rm esc} (r)$ in Cases 2 and 3, respectively. 

\subsection{Simulations}\label{sim}

We performed axisymmetric, relativistic hydrodynamical (rel. HD) simulations in 2 dimensions of the interaction between the jet and the super-Eddington wind. The jet thermal pressure is assumed to be $1$\% of the jet ram pressure divided by $\Gamma_{\rm jet}^2$, so the jet is very cold. The jet streamlines are radial, and the jet has a half-opening angle of 0.1~rad. To optimize the computational cost, the jet initial radius has been taken to be $500\,r_{\rm g}$. This value corresponds to $\approx 7$ cells. This is larger than the usual $\sim 10-100\,r_{\rm g}$ \citep[see e.g.][]{bosch-ramon2006}, but given that most of the wind mass, momentum and energy rates come from the larger radii, the results are not significantly affected. We remark that such an initial jet radius implicitly implies that the actual jet injection height, i.e. the distance from the disk mid-plane, must be larger than the radius. This means that the computational grid does not encompass the jet and wind formation zones, but it is assumed to capture the physics of these flows further downstream. In the perpendicular direction, however, the simulations start from the axis.\par

The wind is launched in the vertical direction for $r>500\,r_{\rm g}$. For $r>r_{\rm cr}$, the wind is injected as a flow of the same velocity than at $r\le r_{\rm cr}$, but with a much lower normalization for the density ($\times 10^{-4}$) and the pressure ($\times 10^{-6}$). Note that in a more realistic setup, instead of being purely vertical, the wind velocity field might start perpendicularly to the surface of the disk. In addition, the wind would carry angular momentum, which should also affect the evolution of the wind on larger scales. Here, however, we want to focus on the impact of the jet on the wind, and on the possibility of formation of an equatorial wind component. Thus, the cleanest way to make such a study as a first approximation is to assume a purely vertical wind.\par

The jet and wind total power, momentum and density rates adopted in theses simulations are those of Cases~1, 2, and 3, given in Table~\ref{table:Parameters}.\par

The code that solves the rel. HD equations is the same as in \citet{delacita2016}: third order in space; second order in time; and using the Marquina flux formula \citep{donat1996,donat1998}. Details on the fluid equations, the implementation of the spatial order scheme, and on the Riemman solver employed can be found in \cite{mignone2005}, \cite{donat1996} and \cite{donat1998}.\par

For the adiabatic index of the gas, we adopted a mono-atomic relativistic ideal gas value of $\gamma=4/3$, which corresponds to a suitable for the relativistic jet, and for an optically thick wind. This approximation fails in the region where the wind becomes optically thin, but for simplicity, we have kept that adiabatic index constant in the whole computational domain. We note that the fact that the wind is optically thick formally prevents efficient cooling, effectively implying that the wind is quasi-adiabatic with $\gamma\approx 4/3$. On the other hand, we assume that the jet is adiabatic, something that is the case under the densities and temperatures adopted in the simulation\footnote{One can still assume that some fraction of the jet energy can be in the form of radiating non-thermal particles, but we assume that such a component will not affect the jet dynamics.}. On larger scales, efficient cooling in an optically thin wind may lead to clumping, instabilities and even shocks, but the study of these effects is beyond the scope of this work.\par

For all cases investigated, the adopted computational grid consist of a uniform grid with 275 cells between $r=z=0$ and $r=z\approx 10^{11}$~cm ($\approx 2\times 10^4\,r_{\rm g}$ for $M_{\mathrm{BH}}=34\,M_\odot$). An extended grid is add with 125 cells in the $\hat{r}$- and $\hat{z}$-directions reaching $r=z=3\times 10^{11}$~cm ($\approx 6\times 10^4\,r_{\rm g}$ for $M_{\mathrm{BH}}=34\,M_\odot$). To avoid numerical artifacts, the $r$-to-$z$ cell aspect ratio is kept $\lesssim 10$ \citep{perucho2005}. Inflow conditions (the jet and the wind) are imposed at the bottom of the grid, reflection at the axis, and outflow in the remaining grid boundaries.\par

In the course of our study, we ran few tests with shorter simulations with two times more resolution and found that the results were consistent with those of the longer, lower resolution simulations. In any case, the first-approach aim of the simulations justify the modest levels of resolution adopted here. However, such a modest resolution can favor jet-wind lateral momentum transfer through numerical viscosity at the jet base, helping to form the equatorial wind. This probably happens in Case 1: the radial jet flow pushes laterally the wind flow, but numerical viscosity can also enhance this effect. We must nevertheless note that the same effect but with a physical (not numerical) origin may result from instability growth under more realistic conditions. In a much higher resolution simulation (beyond the scope of this work), the injection of small fluctuations, expected in the formation of the jet and the wind, both flows with high Reynolds numbers, should induce Kelvin-Helmholtz instabilities in the jet-wind interface leading to mixing of fluids and an effective viscosity.\par

\subsection{Results}

The simulations were run until they reached the steady state within the computational domain. The time needed to reach the steady state was approximately the time needed by the slowest flow in the simulation to cross the domain.
Different simulated times were needed for the jet-wind system to reach a stationary solution: 1112.3~s, 1686.4~s, and 2523.6~s, for Cases~1, 2, and 3, respectively. The slower the densest parts of the wind were, the longer the simulation was.

\begin{figure}[h!]
\centering
\hspace*{-1cm}  
  \begin{minipage}{0.5\textwidth}
\centering
    \includegraphics[scale=.6]{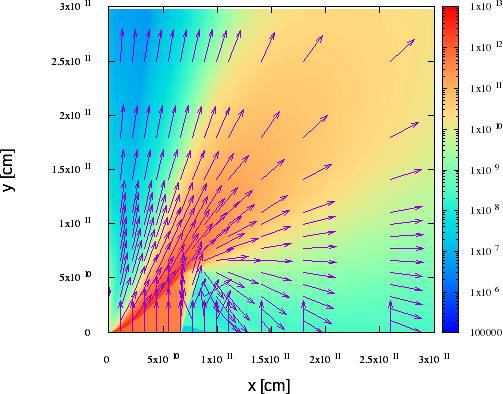}\\
   \end{minipage}%
  \hspace{5mm}
  \begin{minipage}{0.48\textwidth}
    \centering
 \hspace*{-1.cm}  
    \includegraphics[scale=.6]{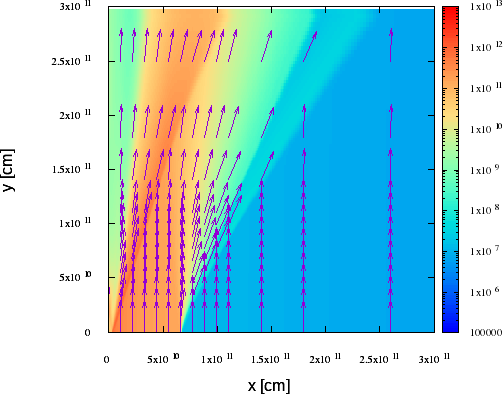}\\
  \end{minipage}
    \hspace{5mm}
  \begin{minipage}{0.48\textwidth}
    \centering
    \hspace*{-1.cm}  
    \includegraphics[scale=.6]{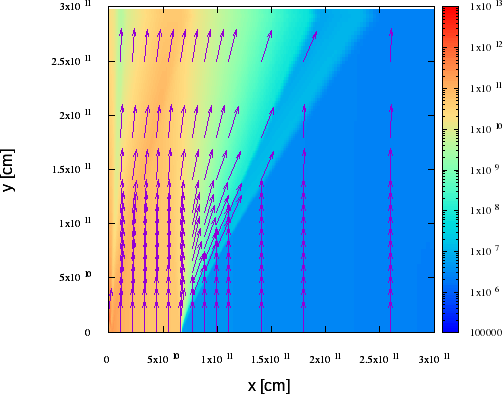}\\
  \end{minipage}
  \caption{Colored density maps (the color scale units are erg~cm$^{-3}$, corresponding to $\rho c^{2}$) of Cases 1 (top), 2 (middle), and 3 (bottom). For the jet and wind parameters, see Table~\ref{table:Parameters}.}
      \label{fig:sim}
\end{figure}

The density maps for all the cases studied are shown in Fig.~\ref{fig:sim}. The results show broader and narrower jets, and winds that are more or less deflected sidewards, depending on the jet-to-wind-power ratio. 

For the case with dominant jet power (Case 1), the jet gets wider than its initial opening angle, getting somewhat recollimated at higher heights, and the wind develops a strong equatorial component. For the intermediate and dominant wind cases (Cases 2 and 3), the jet is well-collimated and the wind keeps mostly vertical.

We have computed the average angle with respect to the $\hat{r}$-direction (i.e. the symmetry plane) at which the wind mass is leaving the grid, and obtained: $44.4^\circ$, $58.2^\circ$, and $58.4^\circ$, for Cases 1, 2, and 3, respectively. Recall that there should be a counter-wind in the $-\hat{z}$-direction, and that lateral jet-wind momentum transfer occurring on larger scales of the structure may lead to somewhat stronger collimation towards the symmetry plane, with the consequent merging of the two wind components in that plane. Thus, it is expected that for Case 1 the wind will look more like a broad equatorial flow than like a vertical one\footnote{The equatorial wind in Population III microquasars does not stop accretion. In these systems the accretion is by overflowing the Roche lobe, therefore the accreted matter flows only in the plane where the Lagrange point is contained, while the equatorial wind is always directed above the accretion disk. On the contrary, accretion arrest could occur if mass transfer from the star is by stellar winds, as in Population I microquasars.}. For Cases 2 and 3, the impact of the jet is smaller; Fig.~\ref{fig:sim} (bottom) basically shows a somewhat widened flow strongly directed upwards. 

In a realistic scenario the jet properties may be different from those found here. We are computing a steady solution of the dynamics of the jet-wind system, although some variability is expected in the injection of both structures, which will surely lead to instability growth downstream the jet. Given that jet variability in super-Eddington accreting sources is beyond the scope of our work, we leave the numerical analysis of such a scenario for future research. In the following sections we consider the effects of a quasi-equatorial wind in PopIII~MQs (Case~1). Since a quasi-equatorial outflow emerges by the influence of the jet on the wind, we refer to it hereafter as the \textit{jet-induced wind}.
\section{Interaction between the jet-induced wind and the donor star}
\label{sect:PopIII}
\subsection{Stagnation of the wind}
From the results obtained in our simulations, we can estimate the velocity of the jet-induced wind at the end of the grid as $v_{\rm w} \approx 3.7\times 10^{8}\,{\rm cm\,s^{-1}}$. We assume that this supersonic velocity is the terminal value.\par

Part of the jet-induced wind flows towards the donor star along the orbit. Since Population III stars have no winds, they are metal-free, the wind can only be stalled by the stellar radiation pressure. The radiation pressure balance the ram pressure of the wind at the point given by:
\begin{equation}
\rho_{\rm w}(r_{\rm \small BH})\,v_{\rm w}^2 = P_{\rm rad} \approx \frac{L_{*}}{12\pi cr_{\rm \small S}^{2}},
\end{equation}
where an optically thick wind is adopted, $r_{\rm \small BH}$ is the distance to the stagnation point measured from the black hole, and $r_{\rm \small S}$ is the distance measured from the star.\par

\begin{figure}[h!]
\centering
\includegraphics[scale=.28]{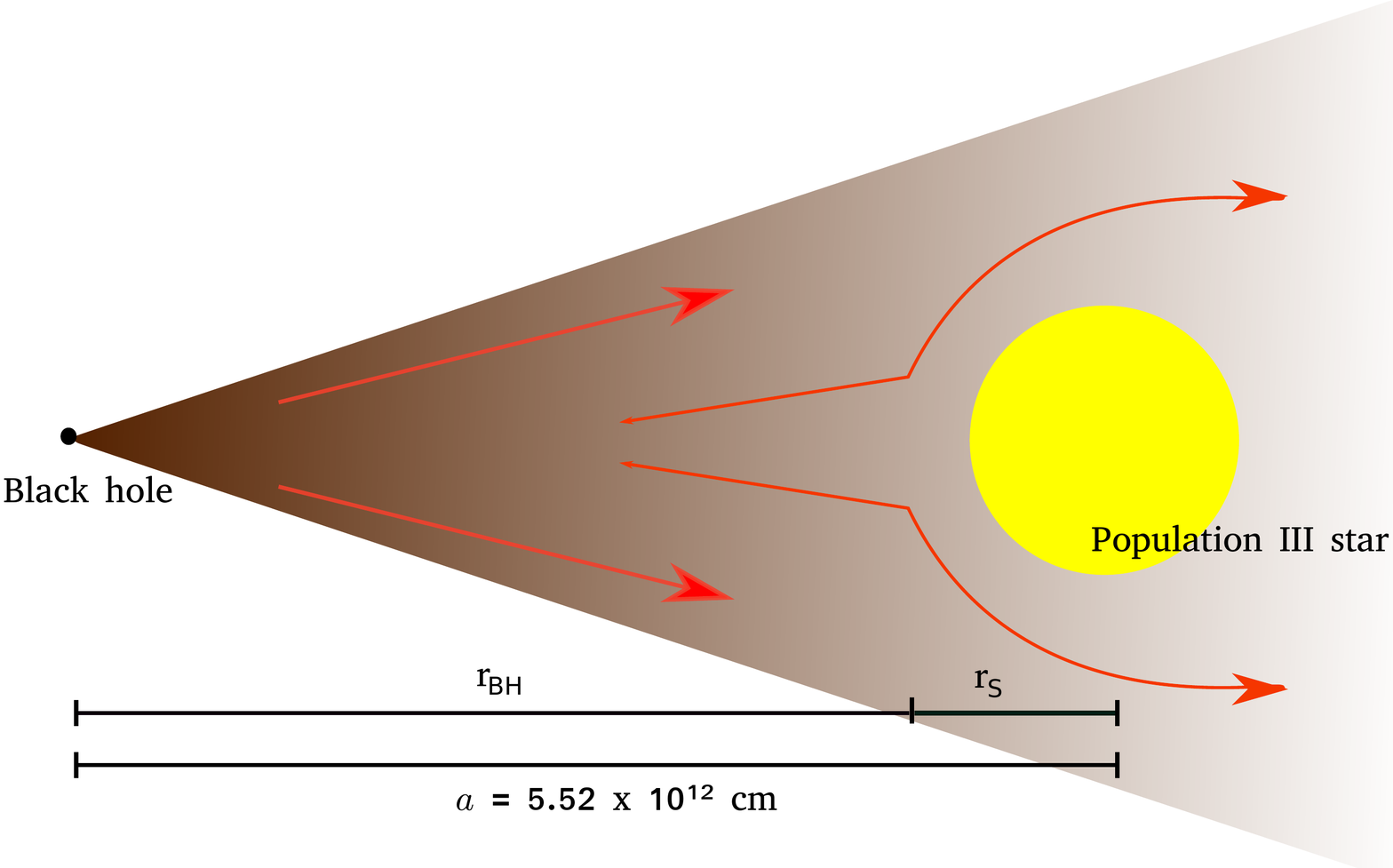}
\caption{Scheme of the interaction between the equatorial wind and the donor star. The wind stagnates at the point where $\rho_{\rm w} v_{\rm w}^2 = P_{\rm rad}$. The distances from the stagnation point to the black hole and the center of the star are indicated as $r_{\rm \small BH}$ and $r_{\rm \small S}$, respectively.}
\label{fig:interaction}
\end{figure}

In the supersonic wind region, the density varies with radius according to $\rho_{\rm w} = \rho_{\rm w,0} (r_{0}/r)^{2}$, where $\rho_{\rm w,0}$ and $r_{0}$ are values at some reference point. We adopt $\rho_{\rm w,0} = 10^{-12}\,{\rm g\,cm^{-3}}$, $r_{0} = 3\times10^{11}\,{\rm cm}$ (see Case 1 in Fig.\ref{fig:sim}), and $ L_{*} = 3\times10^{6}L_{\odot}$. Replacing, we obtain $r_{\rm \small S} \approx 17\,R_{\odot} \approx 1.2\,R_{*} > R_{*}$, then the jet-induced wind is halted by the radiation pressure of the star. This occurs because the luminosity of the first stars is extremely high. Shocks and the associated radiation are expected from this stagnation region of the wind. A forward shock propagates in the tenuous medium dominated by stellar radiation, and a reverse shock moves back towards the black hole. In what follows, we investigate the possibility of particle acceleration in the reverse shock by diffusive shock acceleration, which requires adiabatic conditions to be fullfiled. A sketch of the interaction between the wind and the donor star is shown
in Fig.\ref{fig:interaction}. \par

\subsection{Particle acceleration}

The parameter that determines whether a shock may be treated as adiabatic is the thermal cooling length of the medium traversed by the shock. If the length scale of the medium is higher than the thermal cooling length, the shock is adiabatic, otherwise it is radiative. A detailed discussion of the nature of shocks and their ability to accelerate particles is given in \cite{muller2020}. The thermal cooling length of the wind is given by \citep{mccray1979}:

\begin{equation}
R_{\Lambda} = \frac{5.9\times 10^{-11} \mu \left(v_{\rm sh}/{\rm km\,s^{-1}} \right)^3}{\left(n_{\rm w}/{\rm cm^{-3}} \right)\left(\Lambda(T) / {\rm erg\,s^{-1}\,cm^{-3}} \right)}\;{\rm cm},
\end{equation}
where $\mu = 0.6$ for a fully ionized plasma, $v_{\rm sh} \approx 4\,v_{\rm w}/3$ is the velocity of the reverse shock, $n_{\rm w}$ is the density of the wind in the pre-shocked region, $T = 20 \mu  \left(v_{\rm sh}/{\rm km\,s^{-1}} \right)^2$, and $\Lambda (T)$ is the cooling function. Then, we obtain $v_{\rm sh} \approx 5000\,{\rm km\,s^{-1}}$, and $T \approx 3\times10^8 \,{\rm K}$. For this temperature value, the cooling function is given by \citep{wolfire2003}:
\begin{equation}
\Lambda (T) = 3\times 10^{-27} T^{0.5},
\end{equation}
so, $\Lambda \approx5.2\times 10^{-23}\,{\rm erg\,s^{-1}\,cm^{-3}}$. Replacing, the thermal cooling length is $R_{\Lambda} \approx 10^{12}\,{\rm cm}$.\par
This length scale is on the order of the size of the jet-induced wind, measured from the stagnation point to the black hole. If we consider that the particle acceleration occurs in a small region around the reverse shock, the adiabaticity condition of the reverse shock is guaranteed. We adopt a thickness of $\Delta x_{\rm acc} = 1.2\times10^{8}\,{\rm cm} \left(\approx 3\times 10^{-5}\,r_{\rm \small BH} \right)$ for the particle acceleration zone, so that the ambient fields can be assumed to be homogeneous. Accordingly, we calculate the acceleration and cooling rates of relativistic particles in a stationary one-zone model, for scenarios of efficient and inefficient acceleration in the reverse shock. The adopted parameters are listed in Table \ref{table:wind parameters}.\par

\begin{table*}[t]
\centering \caption{\small Parameters of the particle acceleration region close to the stagnation point of the equatorial outflow.}             
\label{table:wind parameters}      
\begin{tabular}{@{} ll}        
\hline                 
Parameter [Unit] & Value \\    
\hline                        
   $L_{\mathrm{K}}$: Kinetic power of the jet-induced wind [$\mathrm{erg}\,\mathrm{s}^{-1}$]& $1.7\times 10^{37}$ \\
   $v_{\mathrm{wind}}$: Equatorial wind velocity [${\rm cm\,s^{-1}}$] & $3.7\times 10^{8}$\\
   $\theta$: Wind semi-opening angle & $27$\grad \\
   $q_{\mathrm{rel}}$: Content of relativistic particles & $0.1$\\ 
   $\zeta$ Hadron-to-lepton energy ratio & $1$ \\
   $r_{\rm \small S}$: Acceleration point from the center of the star [${\rm cm}$] & $1.2\times10^{12}$ \\
   $r_{\rm \small BH}$: Acceleration point from the black hole [${\rm cm}$] & $4.3\times10^{12}$ \\
   $B$: Magnetic field at the acceleration point [$\mathrm{G}$] & $60$ \\
   $n_{\mathrm{p}}$: Cold matter density at the acceleration point [${\rm cm^{-3}}$] &  $6.7\times 10^{10}$\\
   $\Delta x_{\mathrm{acc}}$: Size of the acceleration region [${\rm cm}$] & $1.2\times 10^{8}$ \\  
   $p$: Injection spectral index & $2.2$\\
   $\eta_{\rm acc}$: Acceleration efficiency & $10^{-4}\,-\,10^{-1}$ \\
   $E_{\mathrm{e}}^{\mathrm{min}}$: Minimum electron energy [$\mathrm{MeV}$] & $1$  \\
\hline                                   
\end{tabular}

\end{table*}

In the acceleration region, a small fraction of the total kinetic power of the jet-induced wind is transferred to relativistic particles by diffusive shock acceleration: $ L_{\mathrm{rel}} = q_{\mathrm{rel}}L_{\mathrm{K}}$. We assume equipartition between relativistic electrons and protons $L_{\mathrm{p}} = \zeta \,L_{\mathrm{e}} = L_{\mathrm{e}}$. The particle injection function is a power-law in the energy $Q(E,z) = Q_{\mathrm{0}}{E^{-p}}$, where $p$ is the spectral index $\sim 2$. The maximum energy $E^{\mathrm{max}}$ that a relativistic particle can attain is obtained by balancing its acceleration and cooling rates (including advection effects).
\par 
The acceleration rate for a charged particle in a magnetic field $B$ is:
\begin{equation}
t_{\mathrm{acc}}^{-1} = \frac{\eta_{\rm acc} e c B}{E},
\end{equation}
where $E$ is the energy of the particle, $\eta_{\rm acc}$ is a parameter that characterizes the efficiency of the acceleration mechanism, and the magnetic field in the acceleration region is given by $B = B_{*}\left(R_{*}/r_{\rm \small S}\right)^{3}$, where $B_*$ is the surface magnetic field of the star\citep[see][]{reimer2006}\footnote{The particle acceleration rate depends on the shock velocity and the diffusion coefficient of the medium. For both shocks parallel or perpendicular to the magnetic field, the diffusion coefficient is a multiple of the Bohm diffusion coefficient, therefore the acceleration timescale is inversely proportional to the gyroradius of the particles \citep[See][]{drury1983}.}. We adopt $B_{*} = 100\,{\rm G}$, similar to the reported for O and B stars in the local Universe \citep{mathys1999}.\par

The total cooling rate is the sum of the radiative cooling rates and adiabatic loss rate $t_{\mathrm{cool}}^{-1} = t_{\mathrm{rad}}^{-1} + t_{\mathrm{ad}}^{-1}$. The adiabatic losses are caused by the expansion of the acceleration region and the radiative losses by the interaction of the relativistic particles with the ambient fields. We consider electron losses by synchrotron radiation, inverse Compton scattering and relativistic Bremsstrahlung, and proton losses by synchrotron radiation, inelastic proton-proton collisions, and photo-hadronic interactions. The target protons for $pp$ collisions and relativistic Bremsstrahlung interactions are the cold protons in the shocked wind, and the target photons for inverse Compton scattering and $p\gamma$ collisions are provided by the radiation field of the star. This field is so strong, that makes all other radiation field sources negligible.
\par
\begin{figure}[h!]
\centering
\includegraphics[scale=0.33]{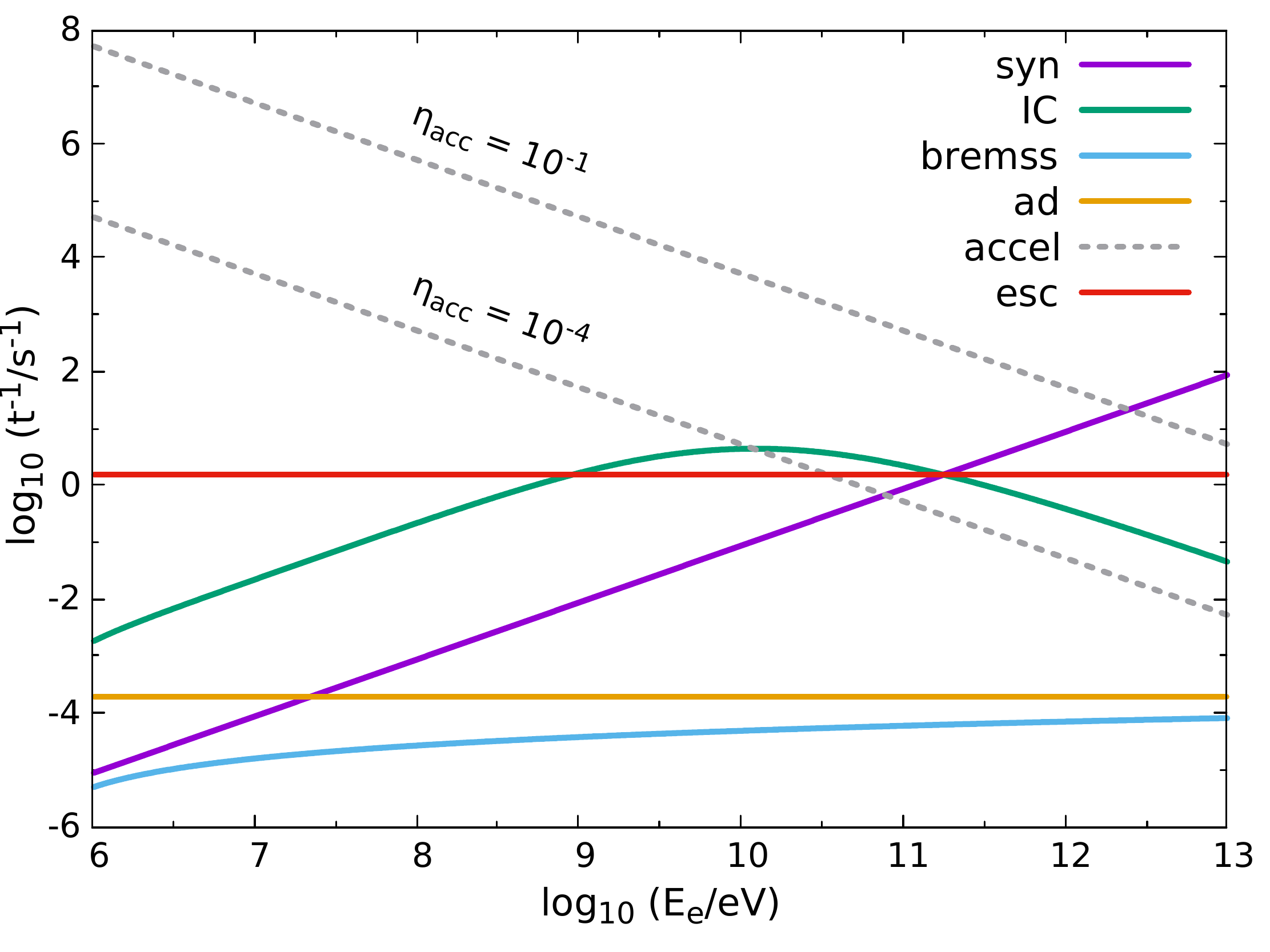}
\caption{\small Rates of acceleration and radiative losses for electrons.}
\label{fig:times electrons}
\end{figure}
\begin{figure}[h!]
\centering
\includegraphics[scale=0.33]{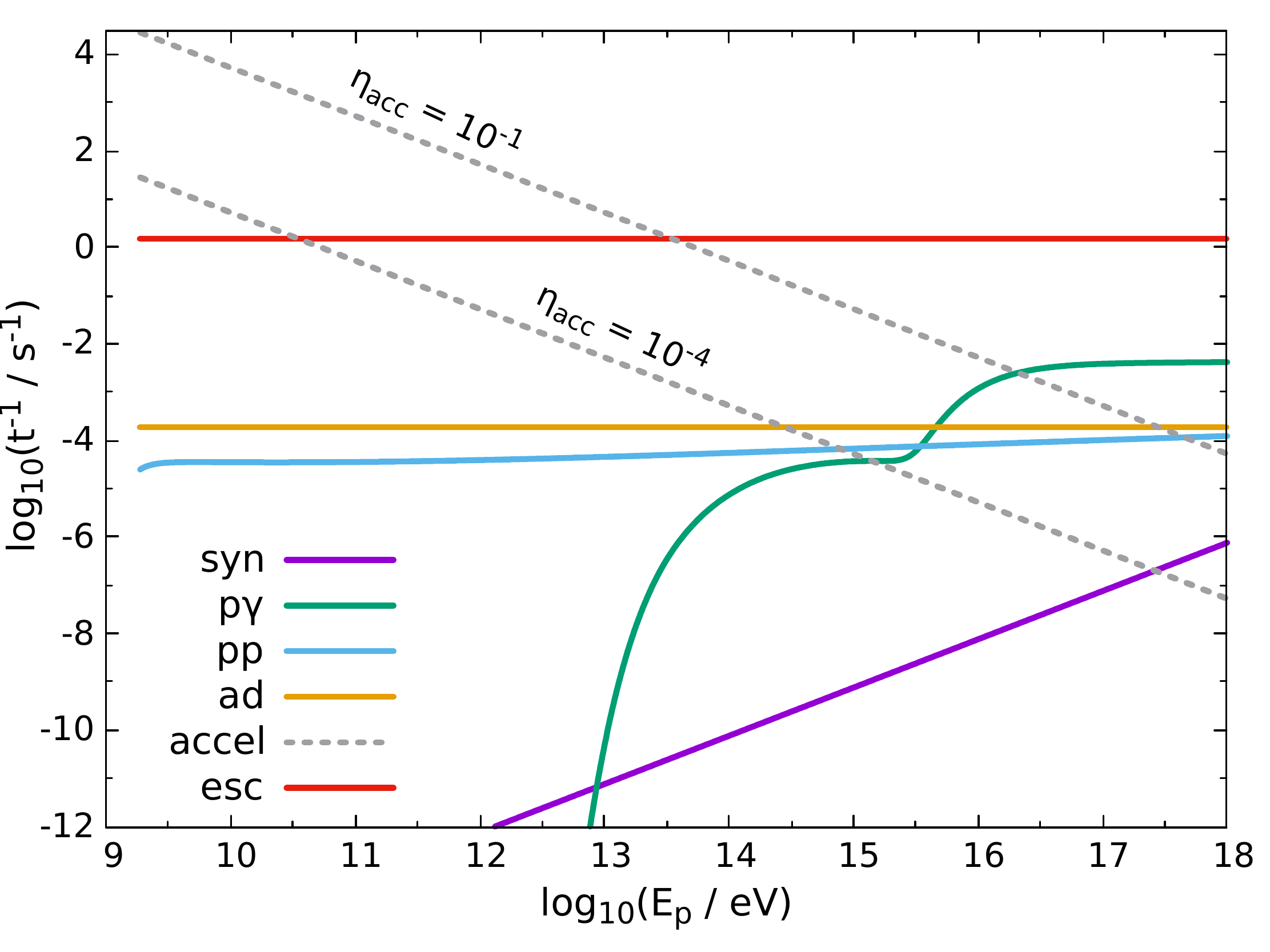}
\caption{\small Rates of acceleration and radiative losses for protons.}
\label{fig:times protons}
\end{figure}

Figures \ref{fig:times electrons} and \ref{fig:times protons} show the cooling and acceleration times in the acceleration region. For $E_{\rm e} < 1\,{\rm GeV}$, the electrons are advected by the shocked wind and escape of the acceleration region without radiative cooling. For $E_{\rm e} > 1\,{\rm GeV}$ electrons cool before escaping by inverse Compton scattering. If the acceleration mechanism is inefficient, the electrons only interact with the photons of the stellar radiation field in the Thomson scattering regime. Otherwise, the electrons reach much higher energies, and are cooled by Compton scattering in the Klein-Nishina regime and synchrotron radiation emission. Protons in any case escape from the acceleration region without cooling, therefore the non-thermal spectrum will be dominated by leptonic processes.
\par
\begin{figure}[h!]
\centering
\includegraphics[scale=0.33]{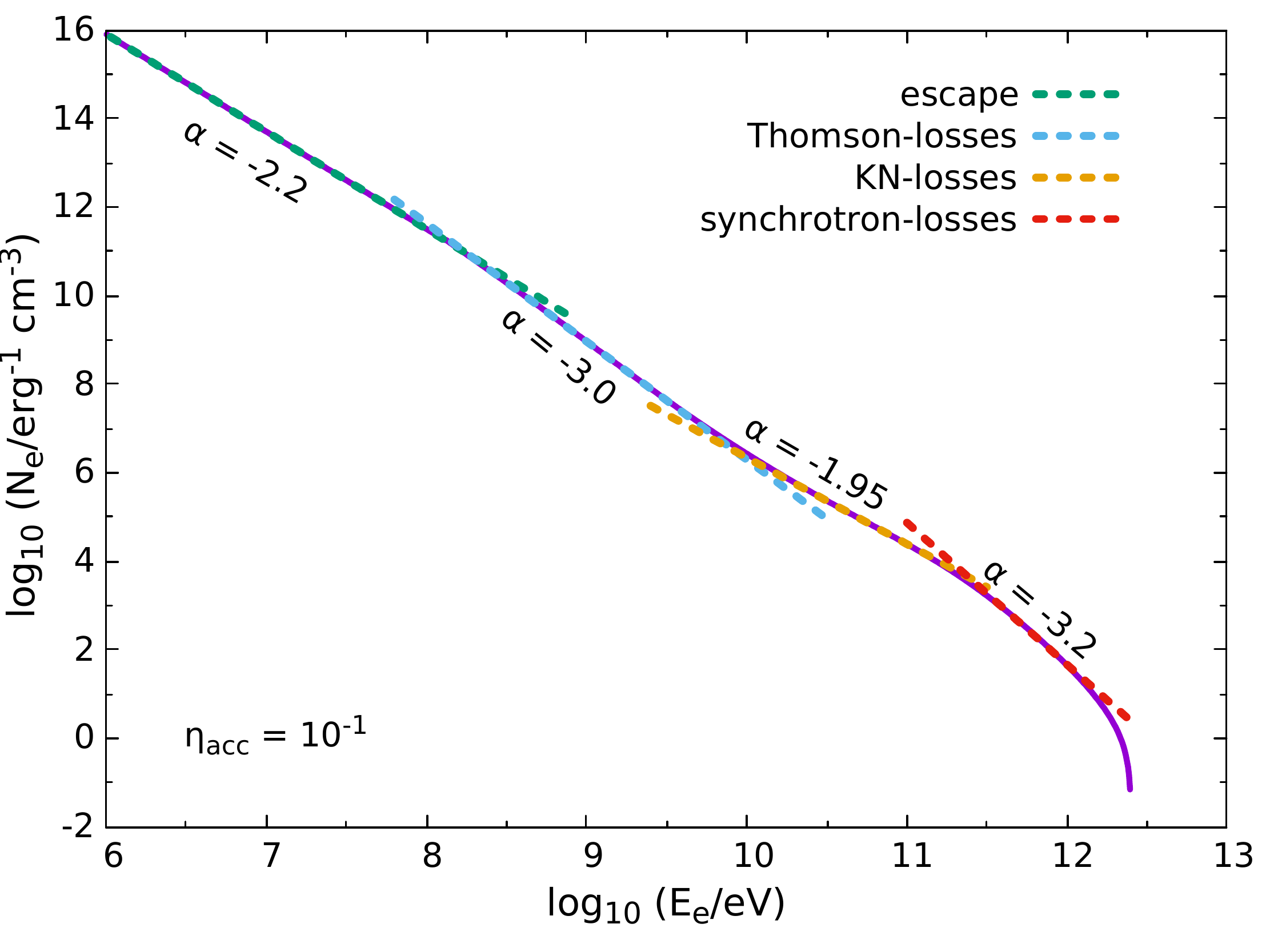}
\caption{Distribution of relativistic electrons in the stationary one-zone model for a highly efficient acceleration.}    
\label{fig:distribution high efficiency}
\end{figure}

\begin{figure}[h!]
\centering
\includegraphics[scale=0.33]{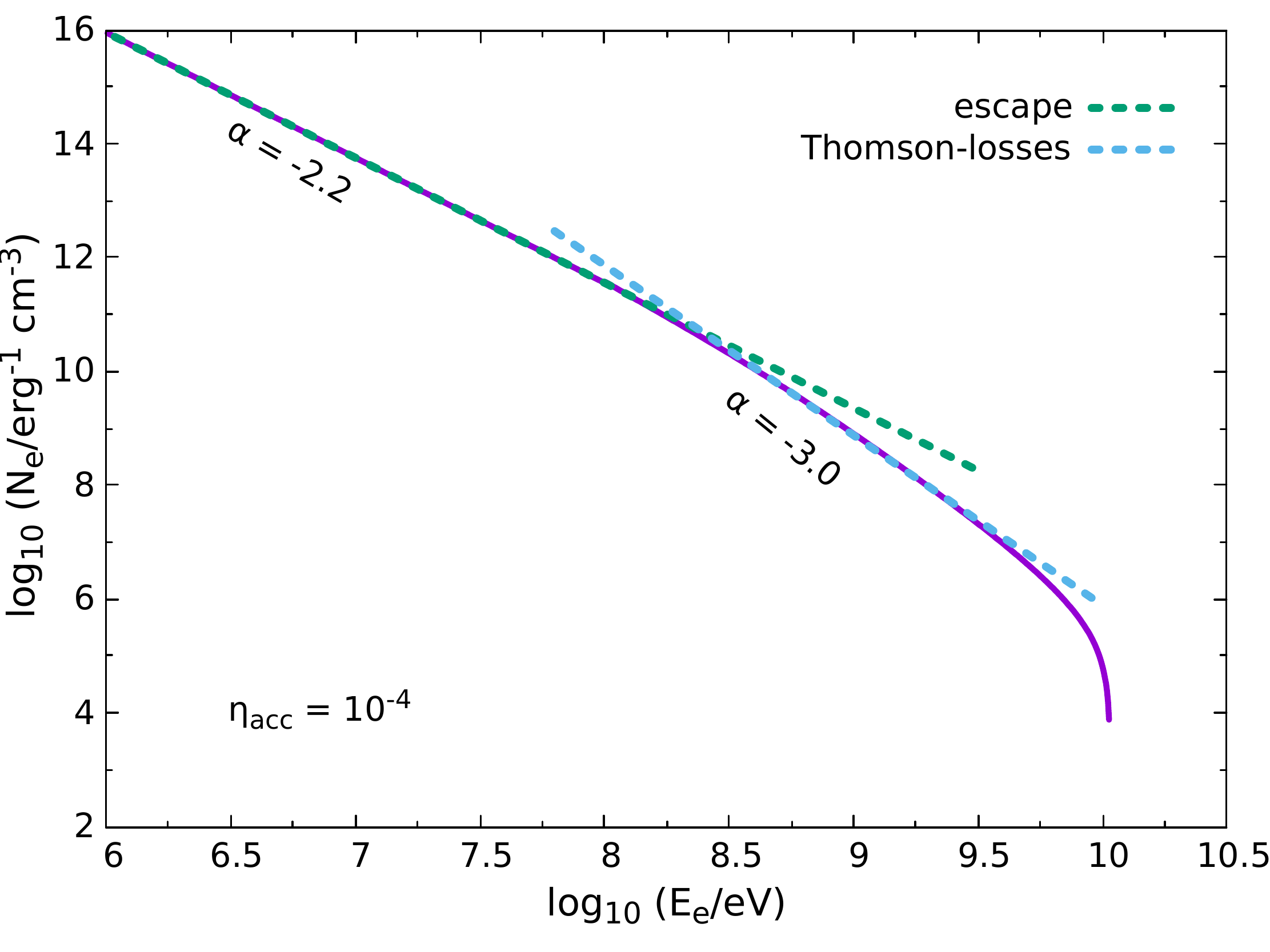}
\caption{Distribution of relativistic electrons in the stationary one-zone model for a low value of acceleration efficiency.}    
\label{fig:distribution low efficiency}
\end{figure}

We solve the transport equation for electrons considering efficient and inefficient acceleration. In Figs. \ref{fig:distribution high efficiency} and \ref{fig:distribution low efficiency} we show the distributions of electrons in steady state in the one-zone model. At low energies, the spectral index of the distribution does not differ from that of the injection function because particle escape dominates. When the radiative processes significantly cool the particles, the spectral index changes, initially producing a softening in the spectrum (when Thomson scattering dominates), then a hardening (in the Klein-Nishina regime), and finally it softens again (mainly because of synchrotron losses).
\par 
\subsection{Nonthermal emission}
\begin{figure}[h!]
 \centering
 \includegraphics[scale=0.33]{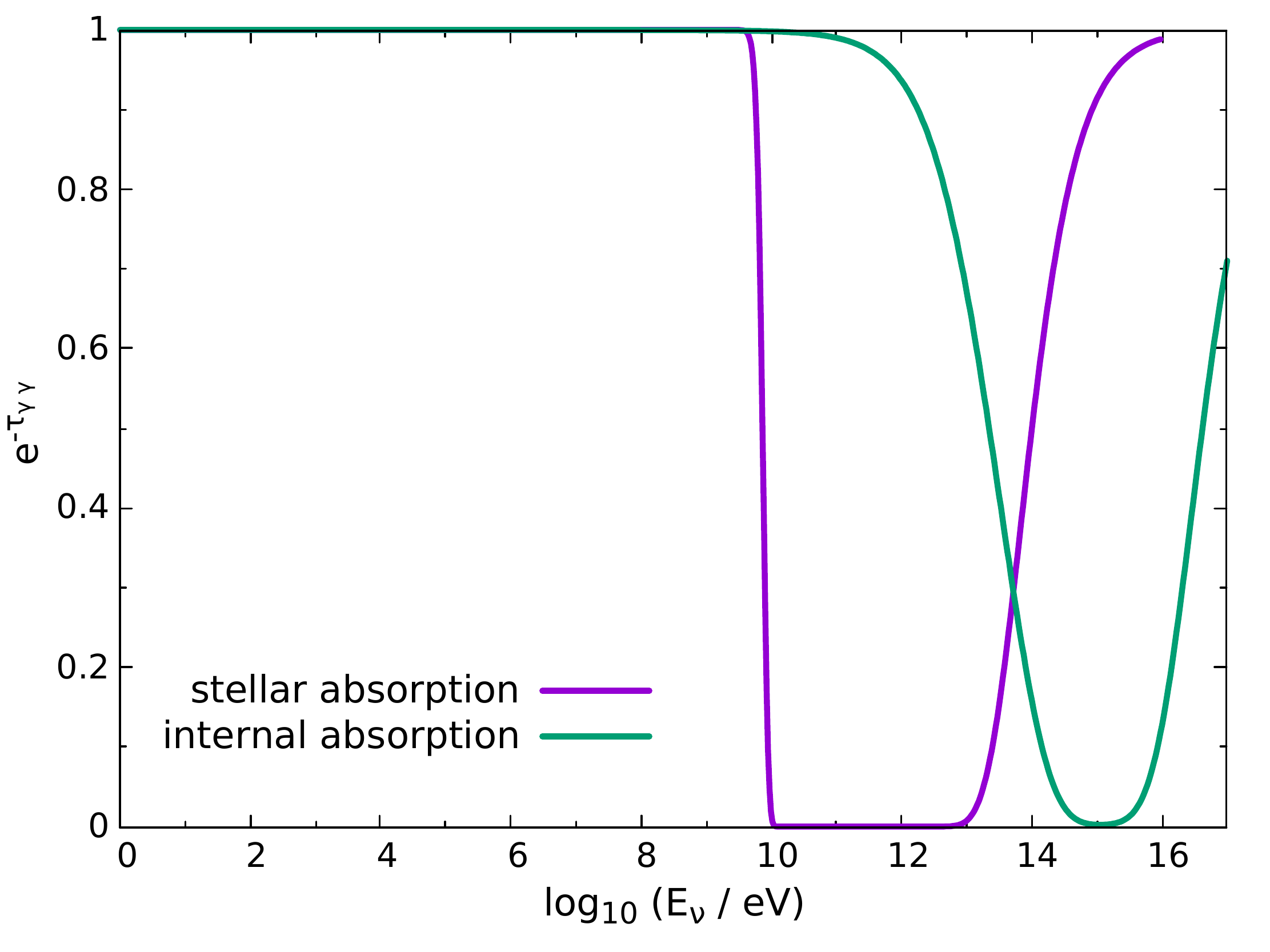}
    \caption{\small Attenuation factor of the emitted radiation by $\gamma \gamma$-annihilation. Internal absorption is caused by interaction with the photons created in the acceleration region by the nonthermal radiative processes. The external stellar radiation field provides the target photons for stellar absorption.}
       \label{fig:attenuation factor}
 \end{figure}

Interactions of the relativistic electrons with the ambient fields give rise to broadband radiation mainly dominated by synchrotron and inverse Compton. The efficiency of the acceleration mechanism determines the high-energy cut-off of the spectrum. Regardless of $\eta_{\rm acc}$, the radiation is totally suppressed for energies $10\,{\rm GeV} < E_{\gamma} < 10\,{\rm TeV}$ by annihilation with photons of the stellar radiation field, as shown in Fig. \ref{fig:attenuation factor}. In addition, internal absorption partially attenuates the emission of $\gamma$ rays for $E_{\gamma} > 1\,{\rm TeV}$. Most of the radiation, however is produced at lower energies.
\par
\begin{figure}[h!]
 \centering
 \includegraphics[scale=0.33]{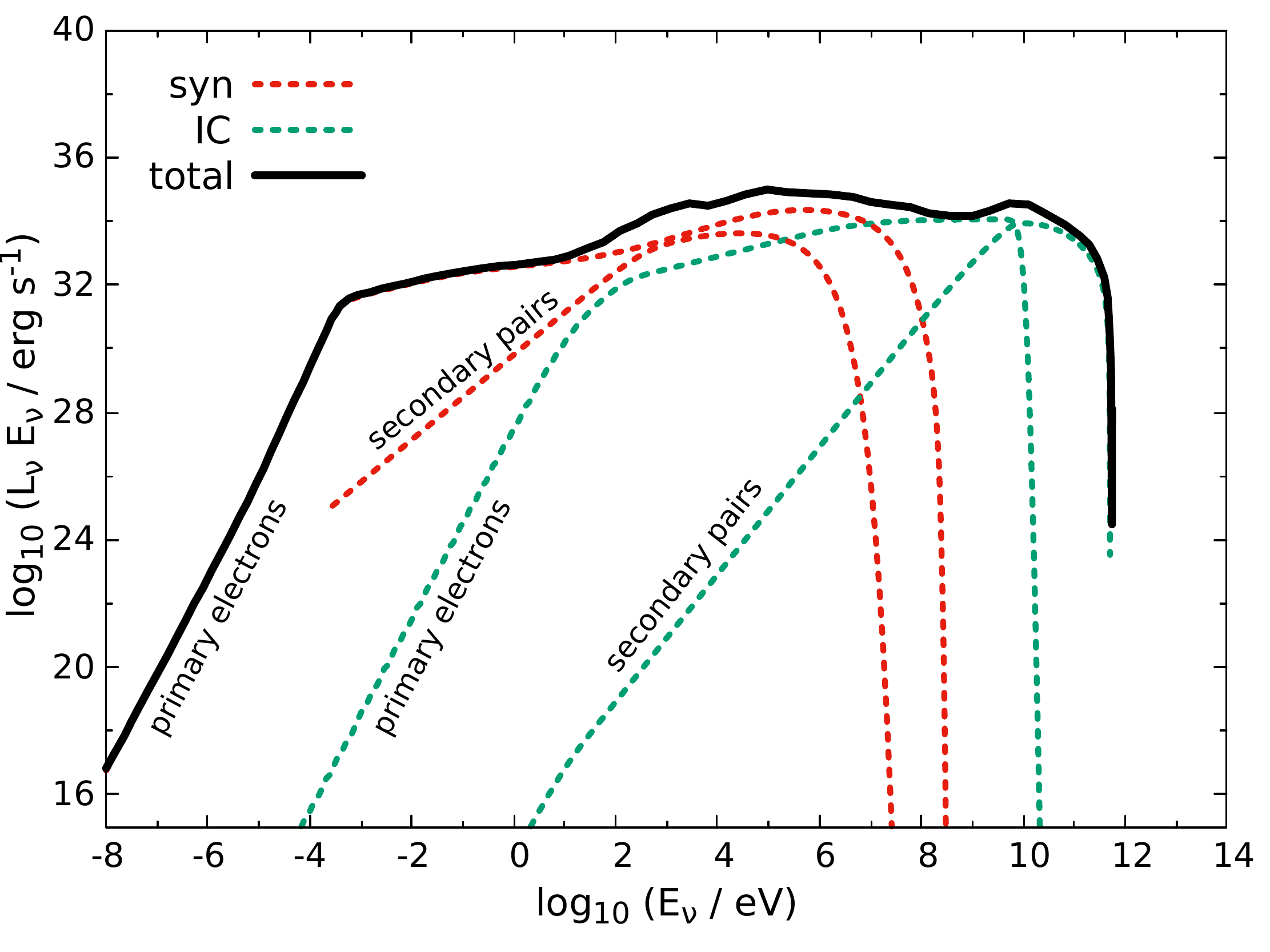}
    \caption{\small Spectral energy distribution of the nonthermal radiation emitted in the acceleration region. Nonthermal radiative processes of the secondary pairs are considered for the calculation of the total spectrum.}
       \label{fig:total SED}
 \end{figure}

The resulting nonthermal spectrum is shown in Fig. \ref{fig:total SED}, including the radiative processes of the first generation of secondary pairs. The absorbed radiation is re-radiated up to energies of 1 TeV by Compton scattering. The processes of $\gamma \gamma$-annihilation, pair creation, and emission of very-high-energy photons by inverse Compton scattering occur only on scales of the binary system.\par
Gamma rays from the inner jet are suppressed for $E_{\nu} > 100\,{\rm MeV}$ by internal absorption \citep{sotomayor2019}. We show here that, nevertheless, photons with energies up to $E_{\gamma} = 1\,{\rm TeV}$ can be expected on binary system scales if the shocks produced by the jet-induced wind can efficiently accelerate particles. Electromagnetic cascades by the interaction of the gamma rays with the cosmic microwave background inject additional high-energy particles in the binary environment.\par
The results shown above do not depend significantly on the size of the acceleration region. In the limit case of considering the entire region between the stagnation point and the surface of the star as the acceleration zone, the proton escape time is still smaller than the radiative cooling time, so a leptonic model is still adequate.\par
For details on the calculation of the cooling rates, emissivities, and opacities, the reader is referred to \cite{bosch-ramon2006,romero2008,romero2010}; and references therein.\par
\section{Summary and conclusions}\label{sect:concl}
We have implemented relativistic hydrodynamical simulations in order to characterize the jet-wind interaction in microquasars with super-Eddington accretion rates. We have applied our study to PopIII~MQs. In order to model the jet-wind interaction in these objects we have adopted the model developed by \cite{sotomayor2019}.\par

We found that the main parameter that determines the jet-wind structure in PopIII~MQs is the relative kinetic power. We have explored three different posibilities: $L_{\rm jet} = 10^{-2}\, L_{\rm wind}$, $L_{\rm jet} = L_{\rm wind}$, and $L_{\rm jet} = 10^{2}\, L_{\rm wind}$. When $L_{\rm jet} << L_{\rm wind}$ (Case 3) or $L_{\rm jet}\sim L_{\rm wind}$ (Case 2), jets keep narrow while the winds do not get significantly deflected sidewards. In such a case, the system is a powerful source of UV radiation and soft X-rays emitted by the wind, and one would expect strong variability of the jet induced by possible wind changes (not explored numerically here). If the kinetic power is similar for both jets and winds, $L_{\rm jet} \approx L_{\rm wind}$, the jet 
may still be somewhat powerful enough to compete with the wind in luminosity, if efficient particle acceleration takes place in the jet. For the case when the kinetic power of the jet is greater than that of the wind, $L_{\rm jet} >> L_{\rm wind}$, the jet strongly affects the structure of the wind, which becomes equatorial and the jet escapes almost without being perturbed. In this case, the jet is the most prominent source of radiation for a distant observer.\par

The geometry we have adopted for the wind of the accretion disk is a purely axisymmetrical flow. In this way, we can analyze more adequately the production of a jet-induced equatorial wind. However, a more realistic geometry should take into account a wind that interacts collisionally with the jet. This would lead to shocks in the jet and to a decrease in jet velocity, likely reducing the capacity of the jet to emerge from the binary system.\par

It is evident that if the luminosity of the Population III star is lower than assumed, the stellar radiation pressure would not halt the equatorial wind. For example, if $L_{*} = 10^{6}L_{\odot}$, then $r_{\rm \small S} \approx 13\,R_{\odot} \approx 0.92\,R_{*}$. In this case, the wind is stopped at the surface of the star. This problem has been studied by \cite{usov1992}, in the context of colliding-wind binaries.\par

When the radiation pressure of the star stagnates the equatorial wind, and a consequent efficient acceleration of particles takes place in the reverse shock, very-high energy gamma rays are injected on scales of the binary system. Electron-positron pairs are created by annihilation, and are expected to deposit their energy at large distances from the binary system. These produce a new reionization channel that should be investigated with numerical simulations.\par

Since in Population III microquasars no appreciable amount of metals is expected in the disk, the radiation drives the wind by Thomson scattering and not by spectral lines as in massive stars in the local Universe. Therefore, we can assume that the wind accelerates only in the lower layer and quickly reaches its terminal velocity, since beyond that the scattered radiation transfers momentum to the wind almost isotropically. This justifies the hydrodynamic treatment at scales of the jet-wind interaction region. Radiation-hydrodynamical simulations are necessary to analyze the same problem in enriched disks.\par

The precession of the jet should increase the transfer of momentum to the wind and modulate it with the precessional phase. A faster wind might be produced that could deposit the bulk of its energy at long distances from the binary system, for example, when interacting with outer dense regions. Our model of jet-induced wind provides a simple mechanism for generating an supersonic equatorial outflow. If this wind is optically thick enough, it can hide the accretion disk and the donor star, as reported in SS433 \citep[see][]{fabrika2004}.\par

Along with wind variability, we have not considered here the effect of baryon loading of the jet by the wind. The simulations consider only one species of particles. However, baryon loading by the wind may be relevant in hadronic microquasars like SS433, where the presence of nuclei in the jets has been revealed through the detection of Fe-lines \citep{migliari2002}. We plan to investigate all these effects and report the results in a future paper.

\acknowledgments
PSC is Fellow of CONICET and PhD student at Universidad Nacional de La Plata (UNLP). PSC thanks the UNLP for the training received during his undergraduate studies, and the public education system in Argentina. GER is very grateful to the ICCUB where part of this research was done. This work was supported by the Argentine agency CONICET (PIP 2014-00338) and the Spanish Ministerio de Ciencia e Innovaci\'on (MICINN) under grant PID2019-105510GBC31 and though the "Center of Excellence Mar\'ia de Maeztu 2020-2023" award to the ICCUB (CEX2019-000918-M). V.B-R. is Correspondent Researcher of CONICET, Argentina, at the IAR.

\bibliographystyle{spr-mp-nameyear-cnd}

\bibliography{biblio-u1}

\end{document}